\documentclass{emulateapj}
\RequirePackage{lineno}
\usepackage{amssymb}
\usepackage{amsmath}
\usepackage{mathrsfs}
\usepackage{natbib}
\usepackage[colorlinks, linkcolor=blue, urlcolor=blue, citecolor=blue]{hyperref}
\usepackage{array}
\usepackage{float}
\usepackage{graphicx}
\usepackage{subfigure}
\usepackage{color}
\usepackage[all]{hypcap}
\usepackage[toc,page]{appendix}

\def\beq{\begin{equation}}
\def\enq{\end{equation}}
\def\bea{\begin{eqnarray}}
\def\ena{\end{eqnarray}}

\begin{document}

\title{Afterglows and Macronovae Associated with Nearby Low-Luminosity Short-Duration Gamma-Ray Bursts}
\author{Di Xiao\altaffilmark{1,2}, Liang-Duan Liu\altaffilmark{1,2}, Zi-Gao Dai\altaffilmark{1,2} and Xue-Feng Wu\altaffilmark{3,4}}
\affil{\altaffilmark{1}School of Astronomy and Space Science, Nanjing University, Nanjing 210093, China; dxiao@nju.edu.cn, dzg@nju.edu.cn}
\affil{\altaffilmark{2}Key Laboratory of Modern Astronomy and Astrophysics (Nanjing University), Ministry of Education, China}
\affil{\altaffilmark{3}Purple Mountain Observatory, Chinese Academy of Sciences, Nanjing 210008, China}
\affil{\altaffilmark{4}Joint Center for Particle, Nuclear Physics and Cosmology, Nanjing
University-Purple Mountain Observatory, Nanjing 210008, China}

\begin{abstract}
A binary neutron star (BNS) merger has been widely argued to be one of the progenitors of a short gamma-ray burst (SGRB). This central engine can be verified if its gravitational-wave (GW) event is detected simultaneously. Once confirmed, this kind of association will be a landmark in multi-messenger astronomy and will greatly enhance our understanding of the BNS merger processes. Due to the limited detection horizon of BNS mergers for the advanced LIGO/Virgo GW observatories, we are inclined to local SGRBs within few hundreds of mega-parsecs. Since normal SGRBs rarely fall into such a close range, to make it more observationally valuable, we have to focus on low-luminosity SGRBs which have a higher statistical occurrence rate and detection probability. However, there is a possibility that an observed low-luminosity SGRB is intrinsically powerful but we are off-axis and only observe its side emission. In this paper, we provide some theoretical predictions of both the off-axis afterglow emission from a nearby SGRB under the assumption of a structured jet and the macronova signal from the ejecta of this GW-detectable BNS merger. From the properties of the afterglow emission, we could distinguish an off-axis normal SGRB from an intrinsically low-energy quasi-isotropic class. Furthermore, with follow-up multi-wavelength observations, a few parameters for BNS mergers (e.g. the medium density and the ejecta mass and velocity) would be constrained.
\end{abstract}


\section{Introduction}
Time domain astronomy has entered a new era since the monumental discovery of gravitational waves (GWs) by the advanced LIGO/Virgo observatories in the last two years \citep{abb16a,abb16b,abb17a,abb17b}. Since then, searching for electromagnetic (EM) counterparts to GWs has become a very urgent issue in this field. Four convinced detections GW 150914, GW 151226, GW 170104 and GW 170814 are believed to originate from binary black hole (BBH) mergers with dozens of solar masses \citep{abb16a,abb16b,abb17a,abb17b}. However, usually we would not expect any EM counterpart from BBH mergers except for several specific situations \citep{con16, loeb16, per16, yam16, zhang16, demi17}. Differing with BBH mergers, the mergers of binary neutron stars (BNSs) are expected to generate a bunch of EM signals, such as short gamma-ray burst (SGRB) jet emission \citep[e.g.][]{fab06, nak07, gia13, ber14, ruiz16, kath17}, cocoon prompt emission \citep{got17, laz17a, laz17b, nak17}, jet/cocoon afterglow \citep[e.g.][]{got17, lamb17, laz17a, nak17}, and macronova \citep{li98, met10, met12, kas13, hot15, got17, nak17}. Besides, a late-time (year-scaled) radio signal might originate from ejecta-medium interactions as the ejecta enters the Sedov-Taylor phase \citep{nak11}.

Although BNS mergers have been proposed as one of the possible progenitors of SGRBs over three decades \citep{pac86, eich89, nar92, moch93} and there are a few indirect evidences for such a scenario \citep[e.g., for reviews see][]{nak07, ber14}, a conclusive proof remains lacking. The detection of its GW emission will provide a unique way to verify this scenario once we can correlate a GW event with an SGRB. However, the advanced LIGO/Virgo GW detection horizon of BNS mergers is about few hundreds of mega-parsecs \citep{aba10, mar16} and SGRBs rarely fall into this close range. Instead, according to the recent statistic analysis of the luminosity function and burst rate of SGRBs \citep{sun15, ghir16}, nearby low-luminosity SGRBs (with luminosities e.g., $L_{\rm iso}<10^{48}\,\rm erg\,s^{-1}$) may be much more numerous than normal ones and we have a greater chance to detect them. Generally, the production of these low-luminosity SGRBs is assigned to less powerful central engines. Nevertheless, there is an another possibility that we are off-axis and only observe the side emission of a normal SGRB. On one hand, even if the GW emission is detected for a BNS merger, we are very likely misaligned with the axis of the SGRB due to the finite small opening angle of a relativistic jet. In the standard picture, it is not easy for us to detect any off-axis EM signal because of the relativistic beaming effect. Alternatively, the side emission from an off-axis SGRB with a structured jet has been discussed as possible EM counterparts to GWs \citep{kath17} and also several other radiation components besides the jet prompt emission have been proposed as possible counterparts in previous works \citep{got17, lamb17, laz17a, laz17b, jin17}. On the other hand, the probability of observing the side emission is estimated to be much higher than the on-axis jet emission \citep{laz17b}. This kind of side emission should be much fainter than the jet from an observational point of view. Based on this argument, we consider several cases that the observing angle $\theta_{\rm obs}$ varies. For large $\theta_{\rm obs}$, it is possible that an observed low-luminosity SGRB is intrinsically energetic. In this paper, therefore, we carry out calculations of multi-wavelength afterglow emission with different observing angles under the assumption of a universally-structured jet and then make a comparison with that of an intrinsically low-energy quasi-isotropic fireball. Our results show that such two types of model are distinguishable and could be tested by follow-up multi-wavelength observations. Furthermore, we explore the macronova emission from a BNS merger for different ejecta parameters and compare it with the afterglow.

This paper is organized as follows. In Section 2 we introduce the universally-structured jet model and calculate the off-axis afterglow emission. Then, we present the method of calculations for the macronova emission in Section 3. Section 4 shows our results for the structured jet model and gives a comparison with an intrinsically low-energy quasi-isotropic fireball. Lastly, we draw conclusions and provide a summary in Section 5.

\section{Off-axis afterglows}
In this section, we consider a structured jet with a lateral distribution of kinetic energy per solid angle $\varepsilon(\theta)$. This kind of jet may form during the propagation of the jet inside the ejecta, which gives rise to shocks at the jet head \citep{nag14, nak17}. Relativistic shocked jet materials form the inner cocoon, which is wrapped by the outer cocoon composed of mildly-relativistic shocked ejecta \citep{got17, nak17, laz17a, laz17b}. Although there is some mixing between them, the cocoon is far from isotropy \citep{nak17, laz17b}. Thus, the overall uniform jet core plus structured cocoon system can be named as a structured jet, of which the kinetic energy per solid angle is assumed to be \citep{dai01, zhang02, ros02, kum03}
\beq
\varepsilon(\theta)\equiv\frac{dE}{d\Omega}=\begin{cases}
\varepsilon_0 & \text{if} \,\,\theta\leq\theta_c, \\
\varepsilon_0(\theta/\theta_c)^{-k} & \text{if}\,\, \theta_c<\theta<\theta_m,
\end{cases}
\label{eq:Edis}
\enq
where the typical half opening angle of SGRBs $\theta_c\approx 16^{\circ}$ \citep{fong15} and the maximum angle $\theta_m=5\theta_c$ are assumed. The index $k$ can be deduced from the luminosity distribution of local event rate density $\rho_0(>L)$. The local event rate density of SGRBs can be fitted by a power-law $\rho_0(>L)\propto L^{-\lambda}$ with $\lambda\sim0.7$ \citep{sun15}. Since $\rho_0(>L)\propto\Omega(>E)\simeq\pi\theta^2$ for similar durations of prompt emission, we can get $L\propto\theta^{-2/\lambda}$. Thus, $\varepsilon(\theta)=dE/d\Omega\propto dL/d\Omega\propto dL/(\theta d\theta)\propto \theta^{-2-2/\lambda}$. Therefore, $k=2+2/\lambda\simeq4.86$, which is consistent with the statistic analysis of \cite{Pes15}. In this paper, we adopt $k=5$ as a nominal value.

For an off-axis observing angle $\theta_{\rm obs}$, the infinitesimal patch of the emission region at $(r,\theta,\phi)$ makes an angle $\alpha$ with respect to the observer, which is given by \citep{kath17}
\beq
\cos\alpha=\cos\theta_{\rm obs}\cos\theta+\sin\theta_{\rm obs}\sin\theta\cos\phi.
\label{eq:ang}
\enq
Assuming that the jet expands outward in a homogeneous medium with a typical number density $n\sim 10^{-2}\,\rm cm^{-3}$ for SGRBs \citep{fong15}, the self-similar evolution of the bulk Lorentz factor $\Gamma$ can be obtained in the same way as previous works \citep[e.g.][]{bm76, huang99, dai01}.
In this paper, the dynamics of a jet follows \cite{huang00} without considering any lateral expansion of the jet. The radius and the time $t^{\prime}$ in the jet's comoving frame can be expressed by
\bea
\frac{dR}{dt}=\frac{c\beta}{1-\beta\cos\alpha},
\ena
and
\bea
\frac{dt^{\prime}}{dt}=\frac{1}{\Gamma(1-\beta\cos\alpha)},
\label{eq:Rtcom}
\ena
where $\beta\equiv(1-1/\Gamma^2)^{1/2}$.

Now we can calculate the synchrotron radiation of the electrons accelerated by the forward shock. Assuming the electrons have a power-law distribution $dn_e/d\gamma_e\propto \gamma_e^{-p}$, the minimum electron Lorentz factor is then $\gamma_m=[(p-2)/(p-1)]\epsilon_e(m_p/m_e)\Gamma$, where $\epsilon_e$ is a fraction of the post-shock energy density converted to electrons and the spectral index of the electron energy distribution $p=2.3$ is adopted as a nominal value. The cooling Lorentz factor is $\gamma_c=6\pi m_ec/(\sigma_TB^{\prime2}t^{\prime})$, where the magnetic field strength in the shocked medium is given by $B^{\prime}=[32\pi\epsilon_B\Gamma(\Gamma-1)nm_pc^2]^{1/2}$ with $\epsilon_B$ being a fraction of the post-shock energy density converted to a magnetic field. In this paper, we adopt typical equipartition factors $\epsilon_e=0.1$ and $\epsilon_B=0.01$ for SGRBs \citep{fong15}. With these parameters, we can calculate the typical frequency $\nu_m^{\prime}$ and the cooling frequency $\nu_c^{\prime}$. According to the relative values of the two frequencies, the spectrum without synchrotron self absorption (SSA) can be written \citep{sari98}. The SSA frequency $\nu_a^{\prime}$ can be obtained by equaling the blackbody luminosity at the Rayleigh-Jeans end with the synchrotron luminosity. At last, we can write down the complete differential luminosity $dL_{\nu'}^{\prime}/d\Omega'$ in the jet's comoving frame \citep[e.g.][]{dai01, xiao17}.

The observed total flux density of the off-axis afterglow is then given by \citep{dai01, gra02, kath17}
\beq
F_\nu=\int_0^{\theta_m}d\theta\int_0^{2\pi}d\phi\frac{dL_{\nu^{\prime}}^{\prime}/d\Omega'}{4\pi D_L^2\Gamma^3(1-\beta\cos\alpha)^3},
\enq
where $D_L$ is the luminosity distance of the source to an observer.
Similarly, the contribution of a counterjet can be accounted for if we integrate $d\theta$ from $\pi-\theta_m$ to $\pi$. The emission from the counterjet is insignificant at the beginning when $\Gamma$ is large but could show up at later times as the structured jet decelerates to a non-relativistic speed. Note that we should integrate on the equal arrival time surface that is determined by $t=\int(1-\beta\cos\alpha)/(c\beta)dR\equiv\rm constant$ \citep{wax97, pan98, sa98, huang00, mod00}.

\begin{figure}
\begin{center}
\includegraphics[scale=0.47]{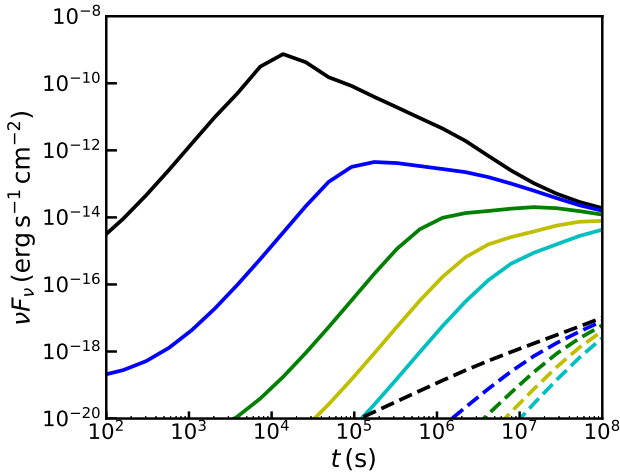}
\caption{The theoretical X-ray light curves for different observing angles. The black, blue, green, yellow and cyan solid lines are corresponding to $\theta_{\rm obs}=0,\, 2\theta_c,\, 3\theta_c,\,4\theta_c,$ and $ 5\theta_c$ respectively. The medium density is taken as $n=10^{-2}\,\rm cm^{-3}$. The corresponding dashed line represent the contribution of a counterjet, which is at least two orders of magnitude fainter.
\label{fig1}}
\end{center}
\end{figure}

\begin{figure}
\begin{center}
\includegraphics[scale=0.45]{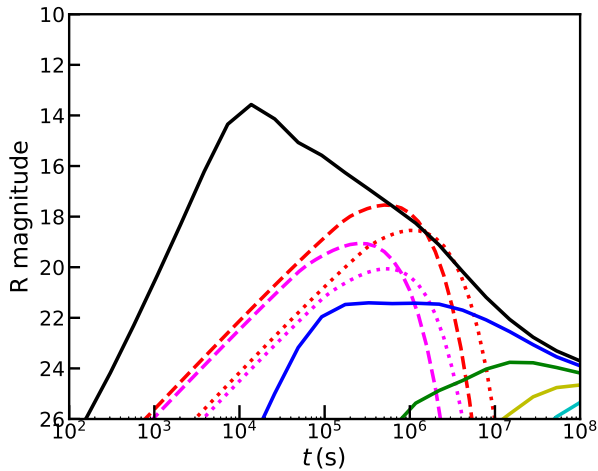}
\caption{The theoretical R-band magnitude for different observing angles. The black, blue, green, yellow and cyan solid lines are corresponding to the afterglow emission of $\theta_{\rm obs}=0,\, 2\theta_c,\, 3\theta_c,\,4\theta_c,$ and $5\theta_c$ respectively. The four macronova signals for $\theta_{\rm ej}=\pi/4$ can be distinguished by colors (magenta for $M_{\rm ej}=10^{-3}M_{\odot}$ and red for $M_{\rm ej}=10^{-2}M_{\odot}$) and line styles (dotted for $v_{\rm ej}=0.1c$ and dashed for $v_{\rm ej}=0.3c$).
\label{fig2}}
\end{center}
\end{figure}

\begin{figure}
\begin{center}
\includegraphics[scale=0.50]{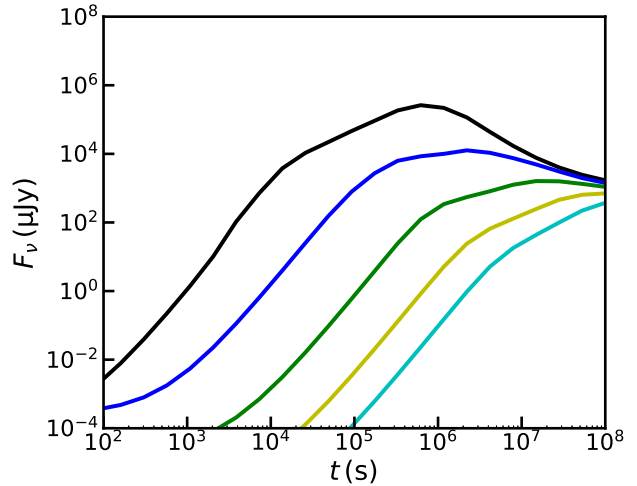}
\caption{The theoretical radio ($\nu=5\,\rm GHz$) light curves for different observing angles. The line styles are the same as in Figure 1.
\label{fig3}}
\end{center}
\end{figure}

\section{Macronovae}
The neutron-rich ejecta produced during the BNS merger
undergoes rapid neutron capture ($r$-process) nucleosynthesis. The
radioactive decay of these heavy nuclei is able to power a day-to-week-long
macronova (also called kilonova)
\citep{li98,kul05,met10,kas13,tan13,met17}.


The density distribution of the ejecta can be obtained from numerical
simulations. The geometry structure of the ejecta can be modeled as a partial sphere in the
latitudinal and longitudinal direction \citep{kyu13,kyu15}.  We
assume a homologous expansion inside the ejecta, so the
density of the ejecta is \citep{kaw16}
\begin{equation}
\rho \left( v,t\right) =\frac{M_{\text{ej}}}{2\phi _{\text{ej}}\theta _{%
\text{ej}}\left( v_{\max }-v_{\min }\right) }v^{-2}t^{-3},
\end{equation}
where $v_{\min }$ and $v_{\text{max}}$ are the minimum and
maximum velocities of the ejecta respectively, $\theta _{\text{ej}}$
is the polar opening angle, and $\phi _{\text{ej}}$ is the azimuthal
opening angle. Here, we adopt $v_{\min }=0.02c,$ and $%
v_{\text{max}}=2v_{\text{ej}}-v_{\min }$. For BNS mergers, there exists a
linearly correlation between $\theta _{\text{ej}}$ and $\phi
_{\text{ej}}$ \citep{die17},
\begin{equation}
\phi _{\text{ej}}=4\theta _{\text{ej}}+\frac{\pi }{2}.
\end{equation}

We assume that the macronova is powered by radioactive decay without an additional energetic engine such as a stable millisecond magnetar.
An approximate expression for the heating rate of $r$-process ejecta is \citep{kor12}
\begin{equation}
\dot{Q}=M_{\text{ej}}\epsilon _{0}\left( \frac{1}{2}-\frac{1}{\pi
}\arctan \frac{t-t_{0}}{\sigma }\right) ^{1.3}\epsilon _{\text{th}},
\end{equation}%
where $\epsilon _{0}=4\times 10^{18}\,\rm erg\, s^{-1}\,{\rm g}^{-1}$, $t_{0}=1.3$ s, and $%
\sigma =0.11$ s are constants. $\epsilon _{\text{th}}$ is the
thermalization efficiency that can be approximated by the fitting
formula \citep{bar16}
\begin{equation}
\epsilon _{\text{th}}=0.36\left[ \exp \left( -a_{1}t_{\text{day}}\right) +%
\frac{\ln \left( 1+2a_{2}t_{\text{day}}^{a_{3}}\right) }{1+2a_{2}t_{\text{day%
}}^{a_{3}}}\right] ,
\end{equation}
where $t_{\text{day}}=t/1$ day, and $a_{1},a_{2}$, and $a_{3}$ are
fitting constants. Here we adopt $a_{1}=0.56,a_{2}=0.17,$ and
$a_{3}=0.74.$

The bolometric luminosity of macronova is approximated by
\citep{kaw16,die17}
\begin{equation}
L_{\text{MN}}=\left( 1+\theta _{\text{ej}}\right) \dot{Q}\times
\left\{
\begin{array}{ll}
t/t_{c}, & {\rm if}\,\,t\leq t_{c}, \\
1, & {\rm if}\,\,t>t_{c},%
\end{array}%
\right.
\end{equation}
The factor $\left( 1+\theta _{\text{ej}}\right) $ indicates the
contribution from radial edge effectively. The critical time $t_{c}$
at which the expanding ejecta becomes optically thin \citep{kaw16} is
\begin{equation}
t_{c}=\left[ \frac{\theta _{\text{ej}}\kappa M_{\text{ej}}}{2\phi _{\text{ej}%
}\left( v_{\max }-v_{\min }\right) c}\right] ^{1/2}.
\end{equation}
For $t<t_{c}$, the mass of the photon-escaping region is
$M_{\text{obs}}(t)=M_{\text{ej}}(t/t_c)$. At $t=t_c$, the whole
region of the ejecta becomes transparent. \cite{kas13} and
\cite{bar13} found that the opacity of $r$-process ejecta,
particularly the lanthanides, is much higher than that for Fe-peak
elements, with $\kappa \sim 10-100$ cm$^{2}$ g$^{-1}.$ In our
analytic model, we adopt $\kappa =10$ cm$^{2}$ g$^{-1}.$
\cite{kaw16} and \cite{die17} found that the bolometric light curve
of the analytic model can well match the results of the
radiation-transfer simulation performed in \cite{tana13}.

Assuming that the macronova is due to blackbody radiation from the photosphere of the ejecta, the effective
temperature can
be written as%
\begin{equation}
T_{\text{eff}}=\left( \frac{L_{\text{MN}}}{\sigma
_{\text{SB}}S}\right) ^{1/4},
\end{equation}%
where $\sigma_{\rm SB}$ is the Stephan-Boltzmann constant and $S=R_{\rm ej}^2\phi_{\rm ej}$
is the emitting area with $R_{\rm ej}\simeq v_{\text{max}}t$ being the radius of the latitudinal edge. The observed flux at
photon frequency $\nu $ can be calculated by
\begin{equation}
F_{\nu ,\text{MN}}=\frac{2\pi h\nu ^{3}}{c^{2}}\frac{1}{\exp \left( h\nu /k_{%
\text{B}}T_{\text{eff}}\right) -1}\frac{R_{\rm ej}^{2}}{D_{L}^{2}},
\end{equation}
where $h$ is the Planck constant and $k_{\rm B}$ is the Boltzmann
constant.

\section{Results}
Figure 1 shows our theoretical X-ray light curves for different observing angles. We consider a typical SGRB with jet core energy $\varepsilon_0=10^{50}\,\rm erg/sterad$, located at a close distance (e.g., $D_L=40\,\rm Mpc$). With the increase of the observing angle, the light curve shifts to later times and the peak luminosity decays, which is consistent with previous works \citep[e.g.][]{mod00,huang00,gra02,lamb17}. For the parameters taken, the contribution from the counterjet (shown by the dashed lines in Figure 1) is negligible.

The light curves of R-band are shown in Figure 2. Different colors represent different observing angles, solid lines are corresponding to afterglow emission, and dashed and dotted lines to macronova emission. The theoretical flux of the macronova signal depends on the kinetic energy and velocity of the ejecta. Numerical simulations have suggested that the ejecta has typical mass $10^{-4}-10^{-2}\,M_{\odot}$ and velocity $0.1-0.3c$ \citep[e.g.][]{nag14}. Thus we consider two masses $10^{-3}M_{\odot}$ (magenta) and $10^{-2}M_{\odot}$ (red), and velocities $0.1c$ (dotted) and $0.3c$ (dashed), so we have four combinations. For large observing angles, the macronova signal might dominate over the afterglow. Therefore, if the macronova component can be extracted in optical-infrared follow-up observations, it will help constrain parameters such as the observing angle and the ejecta mass and velocity.

For completeness, we plot the light curves of the radio band ($\nu=5\,\rm GHz$) in Figure 3. In our model, since the outer structured jet (including the cocoon) first sweeps up the ambient medium, the inner ejecta cannot catch up with the outer jet and thus the ejecta expands freely with a nearly constant velocity. Thus, we neglect any emission from interactions of the ejecta with its ambient gas \citep{nak11}. 

The time evolution of the afterglow spectrum is shown in Figure 4 for the $\theta_{\rm obs}=4\theta_c$ case. Figure 5 shows the influence of the medium density $n$. As is expected, the flux level drops with the decrease of $n$.

\begin{figure}
\begin{center}
\includegraphics[scale=0.47]{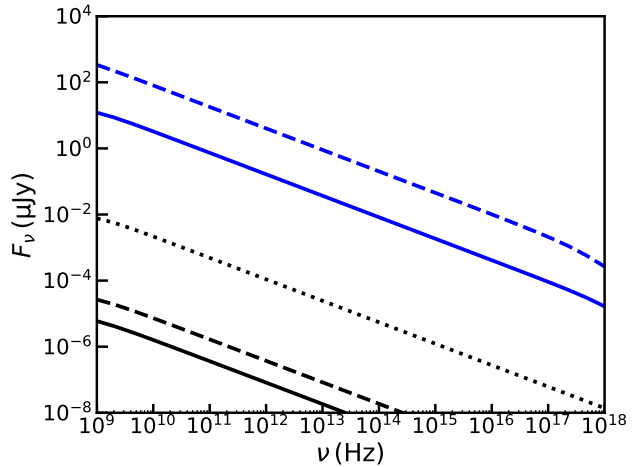}
\caption{The spectrum evolution for the observing angle $\theta_{\rm obs}=4\theta_c$ case. The black solid, dashed, dotted, blue solid, and blue dashed lines represent the spectra at $t=10^3,\,10^4,\,10^5,\,10^6,$ and $10^7\,\rm s$ respectively.
\label{fig4}}
\end{center}
\end{figure}

\begin{figure}
\begin{center}
\includegraphics[scale=0.50]{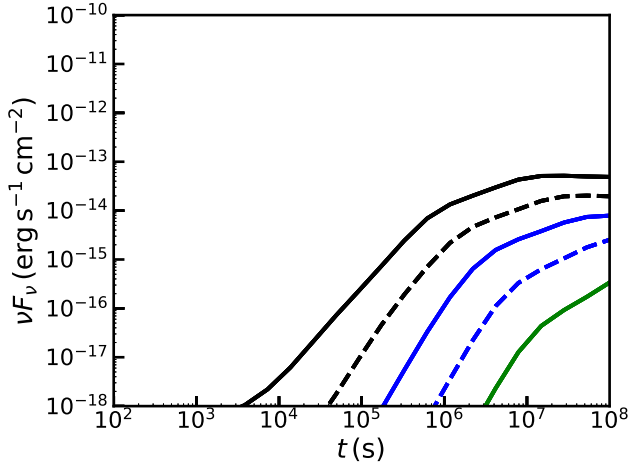}
\caption{The theoretical X-ray light curves for different medium densities for the $\theta_{\rm obs}=4\theta_c$ case. The black solid, dashed, blue solid, dashed and green solid lines are corresponding to $n=1,\,10^{-1},\,10^{-2},\,10^{-3},$ and $10^{-4}\,\rm cm^{-3}$ respectively.
\label{fig5}}
\end{center}
\end{figure}

However, there is still a possibility that an observed low-luminosity burst is not due to the large observing angle, instead, it arises from an intrinsically low-energy quasi-isotropic fireball. We need to consider its afterglow emission for completeness. The structured jet model can be easily generalized to an isotropic fireball case if we set index $k=0$ and opening angle $\theta_m=\pi$ in Equation ({\ref{eq:Edis}). Since the kinetic energy per solid angle along the line of sight in the structured jet model can be estimated by $\varepsilon_0/\varepsilon_{\rm obs}=(\theta_c/\theta_{\rm obs})^{-k}$, to make a direct comparison with one of the previous cases (e.g., $\varepsilon_0=10^{50}\,\rm erg/sterad,\,\theta_{\rm obs}=4\theta_c$), we assume a fireball with isotropic kinetic energy $E_{\rm iso}\sim4\pi\times10^{50}\times4^{-5}\,{\rm erg}\sim1.2\times10^{48}\,{\rm erg}$. The corresponding X-ray, R-band and radio light curves are shown in Figures 6, 7, and 8 respectively. Different lines represent different medium densities, ranging from $n=1-10^{-4}\,\rm cm^{-3}$. We can clearly see that the observed afterglow emission of an intrinsically low-energy fireball is very different with that of an intrinsically powerful off-axis SGRB discussed above. The light curves rise and peak much earlier for the isotropic fireball model. In particular, comparing the blue solid line in Figure 5 with the blue solid line in Figure 6, we can see that the peak time differs to a large extent (almost four orders of magnitude) for these two types of model. Similar differences can be found in R-band ($\sim$ four orders of magnitude) and radio band ($\sim$ three orders of magnitude). Also, the quasi-isotropic macronova signal may be different since intrinsically-fainter SGRBs are likely accompanied by less energetic ejecta, so the macronova should be dimmer. The spectral evolution with time is also different from each other in the two types of model if we compare Figure 9 with Figure 4. All of these results would be testable by follow-up multi-wavelength observations.

\begin{figure}
\begin{center}
\includegraphics[scale=0.47]{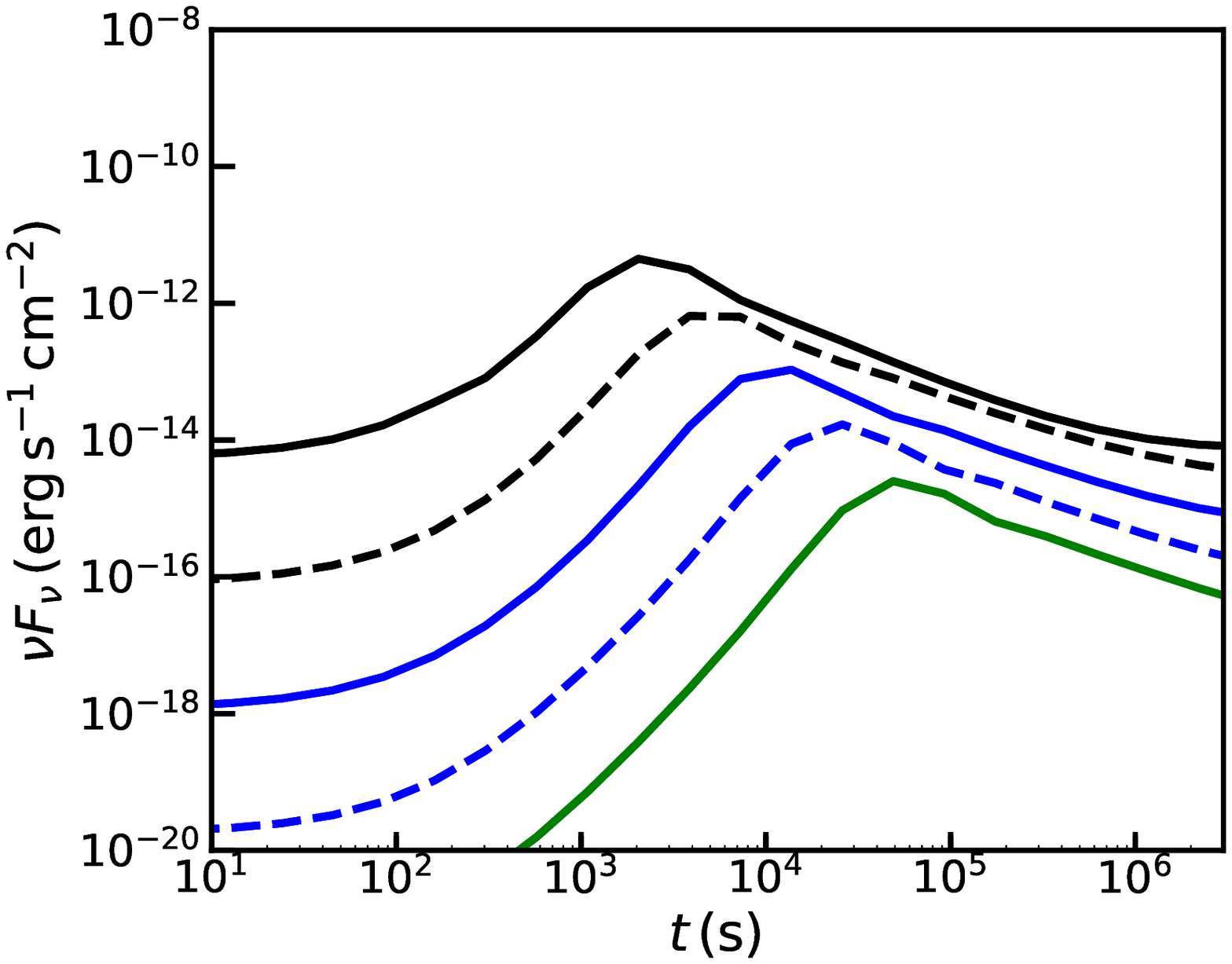}
\caption{The theoretical X-ray light curves for different medium densities for the $E_{\rm iso}=1.2\times10^{48}\,{\rm erg}$ fireball case. The black solid, dashed, blue solid, dashed and green solid lines are corresponding to $n=1,\,10^{-1},\,10^{-2},\,10^{-3},$ and $10^{-4}\,\rm cm^{-3}$ respectively.
\label{fig6}}
\end{center}
\end{figure}

\begin{figure}
\begin{center}
\includegraphics[scale=0.47]{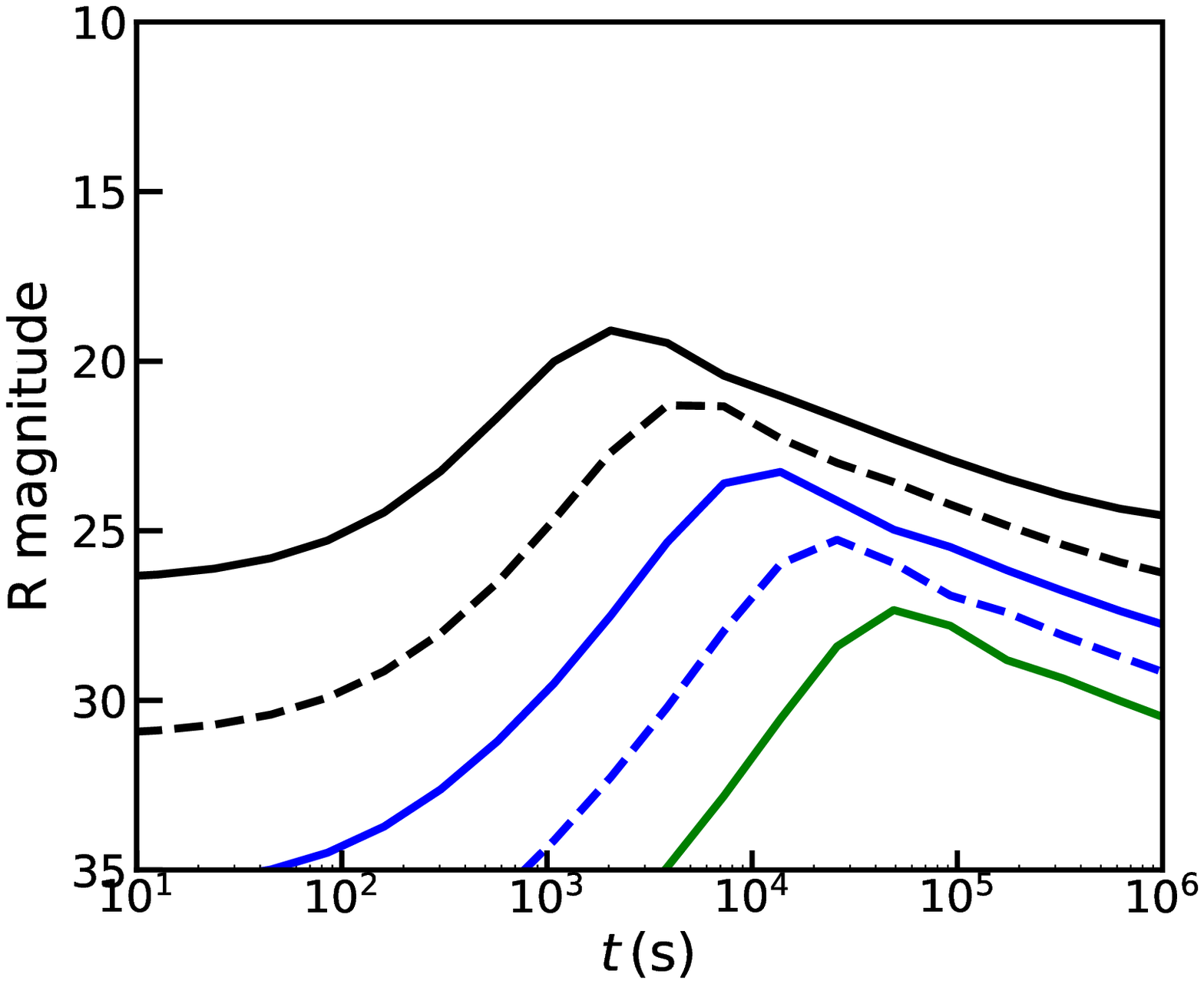}
\caption{The theoretical R-band light curves for different medium densities for the $E_{\rm iso}=1.2\times10^{48}\,{\rm erg}$ fireball case. The line styles are the same as in Figure 6.
\label{fig7}}
\end{center}
\end{figure}

\begin{figure}
\begin{center}
\includegraphics[scale=0.50]{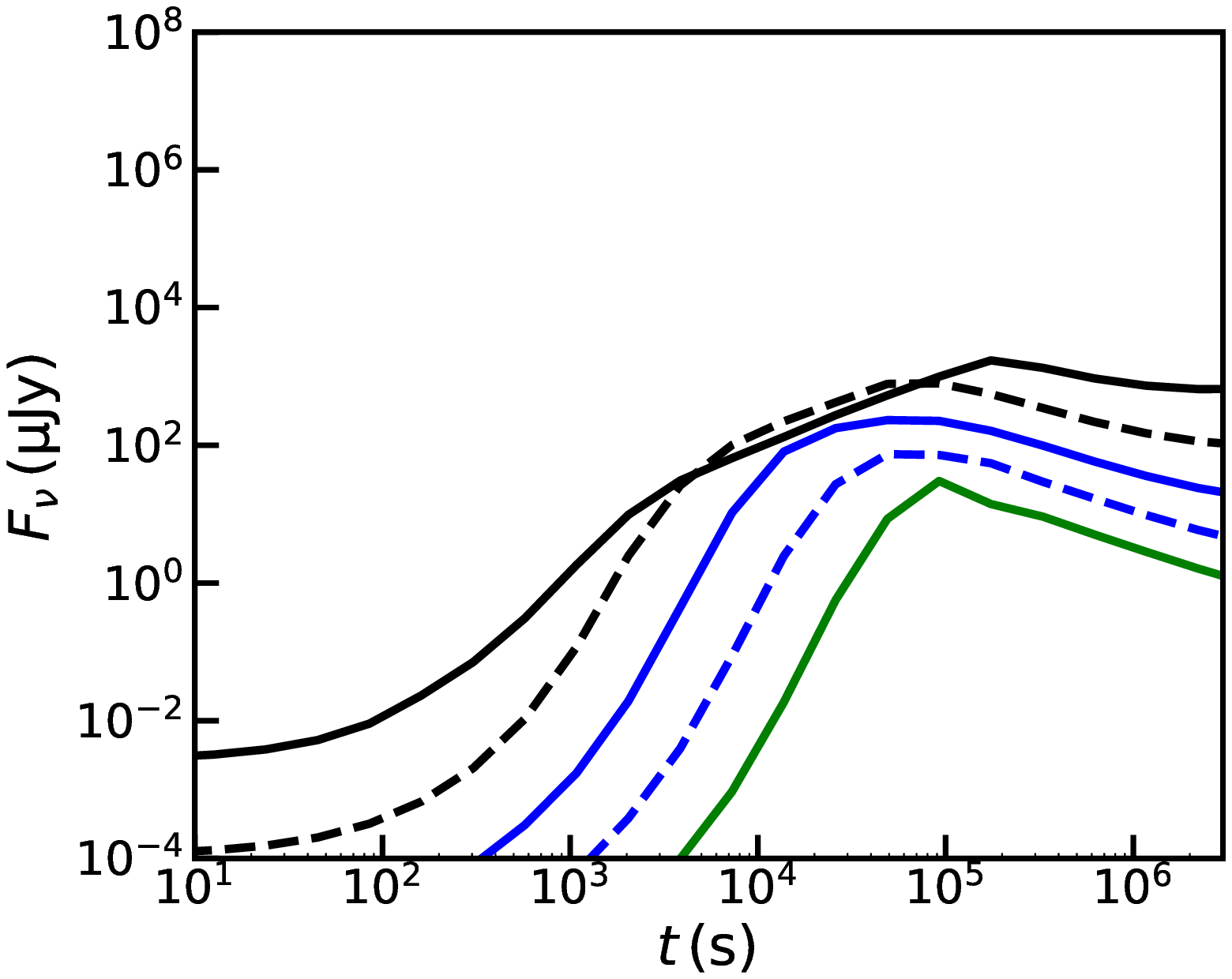}
\caption{The theoretical radio (5\,GHz) light curves for different medium densities for the $E_{\rm iso}=1.2\times10^{48}\,{\rm erg}$ fireball case. The line styles are the same as in Figure 6.
\label{fig8}}
\end{center}
\end{figure}

\begin{figure}
\begin{center}
\includegraphics[scale=0.47]{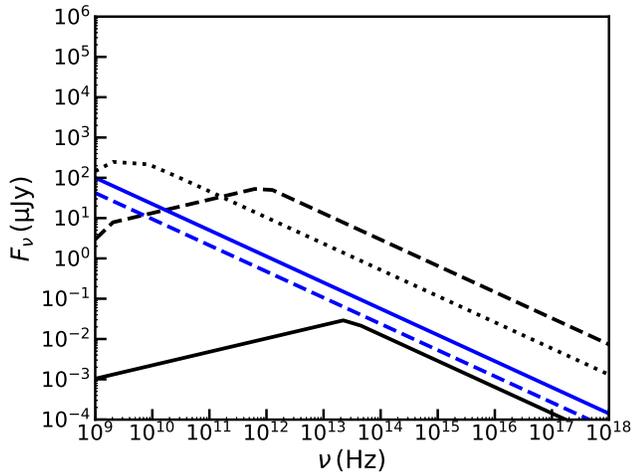}
\caption{The spectrum evolution for the isotropic fireball case. The black solid, dashed, dotted, blue solid, and blue dashed lines represent the spectra at $t=10^3,\,10^4,\,10^5,\,10^6,$ and $10^7\,\rm s$ respectively.
\label{fig9}}
\end{center}
\end{figure}

\section{Conclusions}
\label{sec:disc}
In this work we have re-investigated both an afterglow and a macronova, which are associated with a nearby low-luminosity SGRB from BNS mergers, under the assumption of a universally structured jet. {\em The discovery of such EM signals associated with a GW event in the future will mark the coming of multi-messenger time-domain astronomy}. Detecting a low-luminosity SGRB is estimated to be much easier in our local universe than a normal one because the former has a much greater occurrence rate than the latter does. We have considered two possibilities that either an SGRB is intrinsically low-luminosity and quasi-isotropic or it is just due to off-axis jet emission. We have shown that the properties of afterglow emission in these two cases are obviously different. The light curves rise slower and peak at a later time for the off-axis case. The spectrum is also different at any given time. Furthermore, if we assume the kinetic energy of the ejecta is in proportional to that of the jet, the macronova signal in the low-energy fireball case should be much fainter than that of the off-axis powerful SGRB case. With follow-up multi-wavelength observations of a local low-luminosity SGRB, we can finally distinguish between these two models. In addition, several key parameters can be constrained such as the observing angle, the ejecta mass, the ejecta velocity, and the ambient medium density, all of which would help reveal the mystery of SGRBs. Finally, most importantly, if this kind of low-luminosity SGRB from a BNS merger is associated with a GW event, our theoretical results could provide some hints on searching for EM counterparts to GWs, which would greatly enhance our understanding of the BNS merger processes.

\acknowledgements
This work was supported by the Strategic Priority Research Program ``Multi-waveband gravitational wave Universe'' (grant No. XDB23040000) of the Chinese Academy of Sciences, the National Basic Research Program of China (973 Program grant 2014CB845800) and the National Natural Science Foundation of China grant 11573014.  XFW was also partially supported by the Youth Innovation Promotion Association (No. 2011231), the Key Research Program of Frontier Sciences (QYZDB-SSW-SYS005), and the National Natural Science Foundation of China grant 11673068, 11433009.


\clearpage

\end{document}